\documentclass[showpacs,preprintnumbers,aps,amsmath,amssymb,twocolumn]{revtex4}
\usepackage{graphicx}

\begin{document}

\title{Feigenbaum graphs at the onset of chaos}

\author{Bartolo Luque$^1$, Lucas Lacasa$^1$ and Alberto Robledo$^{2,*}$}

\affiliation{$^1$ Dept. Matem\'{a}tica Aplicada y Estad\'{i}stica. ETSI Aeron\'{a}uticos, Universidad Polit\'{e}cnica de Madrid, Spain.\\
$^2$ Instituto de F\'{\i}sica y Centro de Ciencias de la Complejidad, Universidad Nacional Aut\'{o}noma de M\'{e}xico, Mexico.}
\email{robledo@fisica.unam.mx, tel:+34913366326}


\begin{abstract}

We analyze the properties of networks obtained from the trajectories of unimodal maps at the transition to chaos via the horizontal visibility (HV) algorithm. We find that the network degrees fluctuate at all scales with amplitude that increases as the size of the network grows, and can be described by a spectrum of graph-theoretical generalized Lyapunov exponents. We further define an entropy growth rate that describes the amount of information created along paths in network space, and find that such entropy growth rate coincides with the spectrum of generalized graph-theoretical exponents, constituting a set of Pesin-like identities for the network.

\end{abstract}

\maketitle

\section{Introduction}
Pesin's theorem \cite{dorfman1} prescribes the equality of the
Kolmogorov-Sinai (KS) entropy $h_{KS}$ with the sum of the positive Lyapunov
exponents $\lambda _{i}>0$ of a dynamical system, i.e. $h_{KS}=\sum_{i}%
\lambda _{i}$. The former is a rate of entropy growth that is a metric
invariant of the dynamical system, while the latter is the total asymptotic
expansion rate present in the chaotic dynamics. This relation provides a
deep connection between equilibrium statistical mechanics and chaos. In the
limiting case of vanishing Lyapunov exponent, as in the onset of chaos in
one-dimensional nonlinear maps, Pesin's theorem is still valid but there are
important underlying circumstances that are not expressed by the trivial
identity $h_{KS}=\lambda =0$. At variance with the chaotic region, at the transition
to chaos phase space is no
longer visited in an ergodic way and trajectories within the attractor show
self-similar temporal structures, they preserve memory of their previous
locations and do not have the mixing property of chaotic trajectories \cite%
{baldovin1, mayoral1}. For chaotic and periodic attractors the sensitivity to initial conditions converges to an exponential function for large iteration times giving rise to the familiar positive or negative Lyapunov exponents associated with them. At the transition to chaos exponential separation or merging of trajectories no longer occurs, but notably the sensitivity to initial conditions does not converge to any single-valued function and, on the contrary, displays fluctuations that grow indefinitely with time. For
initial positions on the attractor the sensitivity develops a universal
self-similar temporal structure and its envelope grows with iteration time $%
t $ as a power law \cite{baldovin1, mayoral1}. For unimodal maps it has been
shown \cite{baldovin1, mayoral1} that this rich borderline condition accepts
a description via a spectrum of generalized Lyapunov exponents that matches
a spectrum of generalized entropy growth rates obtained from a scalar
deformation of the ordinary entropy functional, the so-called Tsallis
entropy expression \cite{tsallis1, tsallisbook}. The entropy rate also
differs from the KS entropy in that it is local in iteration time. That is, the entropy rate at time $t$ is determined only by the positions of trajectories at this time while the KS entropy considers all previous sets of positions. Here we
present evidence that this general scenario of unimodal maps at the onset of
chaos is captured by a special type of complex network.

Very recently \cite{tolo1, tolo11}, the horizontal visibility (HV) algorithm
\cite{tolo2, tolo3} that transforms time series into networks has offered a
view of chaos and its genesis in low-dimensional maps from an unusual
perspective favorable for the discovery of novel features and new
understanding. We focus here on networks generated by unimodal maps at their
period-doubling accumulation points and characterize the fluctuations in
connectivity as the network size grows. We show that the expansion of
connectivity fluctuations admits the definition of a graph-theoretical
Lyapunov exponent. Furthermore, the entropic functional that quantifies
the amount of information generated by the expansion rate of trajectories in
the original map appears to translate, in the light of the scaling properties of the
resulting network, into a generalized entropy that surprisingly coincides
with the spectrum of generalized graph-theoretical Lyapunov exponents. This suggests
that Pesin-like identities valid at the onset of chaos could be found in
complex networks that possess certain scaling properties.

\noindent The rest of the Letter is as follows: We first recall the HV algorithm
that converts a time series into a network and focus on the so-called
Feigenbaum graphs \cite{tolo1, tolo11} as the subfamily of HV graphs
generated by iterated nonlinear one-dimensional maps. We then expose the
universal scale-invariant structure of the Feigenbaum graphs that arise at
the period-doubling accumulation points. We subsequently show how to define
a graph-theoretical Lyapunov exponent in this context. In agreement with the
known dynamics of unimodal maps at the transition to chaos the
graph-theoretical Lyapunov exponent vanishes, and proceed to define
generalized exponents that take into account the subexponential expansion of
connectivity fluctuations. We finally show that the Feigenbaum graph that
represents the onset of chaos admits a spectrum of generalized graph-theoretical exponents that coincide with a spectrum of deformed entropies.

\begin{figure}[t]
\centering
\includegraphics[width=0.48\textwidth]{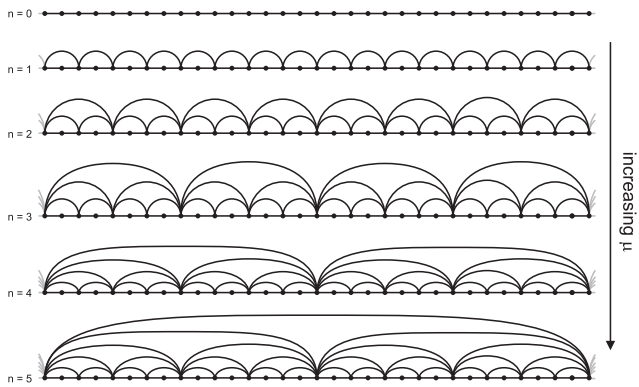}
\caption{{\protect\small {Feigenbaum graphs associated with periodic series of
increasing period $2^{n}$ undergoing a period-doubling cascade. The
resulting patterns follow from the universal order with which an orbit
visits the positions of the attractor. The Feigenbaum graph associated with the time series generated at the onset of chaos ($n\rightarrow \infty$) is the result of an infinite application of the inflationary process by which a graph at period $2^{n+1}$ is generated out of a graph at period $2^{n}$ \cite{tolo1}.}}}
\label{grafos de feigenbaum1}
\end{figure}

\section{The Feigenbaum graph at the onset of chaos}
The horizontal visibility (HV) algorithm is a general method to convert
time series data into a graph \cite{tolo2, tolo3} and is concisely stated as
follows: assign a node $i$ to each datum $x_{i}$ of the time series $%
\{x_{i}\}_{i=1,...,N}$ of $N$\ real data, and then connect any pair of nodes
$i$, $j$ if their associated data fulfill the criterion $x_{i}$, $%
x_{j}>x_{n} $ for all $n$ such that $i<n<j$. The capability of the method as well as other variants to
transfer properties of different types of time series into their resultant
graphs has been demonstrated in recent works \cite{tolo4, toral}. When the
series under study are the trajectories within the attractors generated by
unimodal or circle maps the application of the HV algorithm yield
subfamilies of visibility graphs, named Feigenbaum and Quasiperiodic graphs respectively, that render the
known low-dimensional routes to chaos in a new setting \cite{tolo1, tolo11,
quasi}. For illustrative purposes, in Fig. \ref{grafos de feigenbaum1} we
show a hierarchy of Feigenbaum graphs obtained along the period-doubling
bifurcation cascade of unimodal maps. At the accumulation point of the
cascade the trajectories within its non-chaotic multifractal attractor
become aperiodic, and in analogy with the original Feigenbaum treatment \cite%
{Feigenbaum}, the associated Feigenbaum graphs evidence scaling properties
that can be exploited by an appropriate graph-theoretical Renormalization Group transformation
\cite{tolo1, tolo11}. Because the order of visits of positions of periodic
attractors or of bands of chaotic attractors in unimodal maps are universal,
a relevant consequence of the HV criterion is that the resulting Feigenbaum
graph at the onset of chaos is the same for every unimodal map. That is, it
is independent of the shape and nonlinearity of the map \cite{tolo1, tolo11}.
This permits us to concentrate our study on a specific unimodal map, that
for simplicity will be the logistic map, and claim generality of the results.\\

\begin{figure*}[tbh]
\centering
\includegraphics[width=0.85\textwidth]{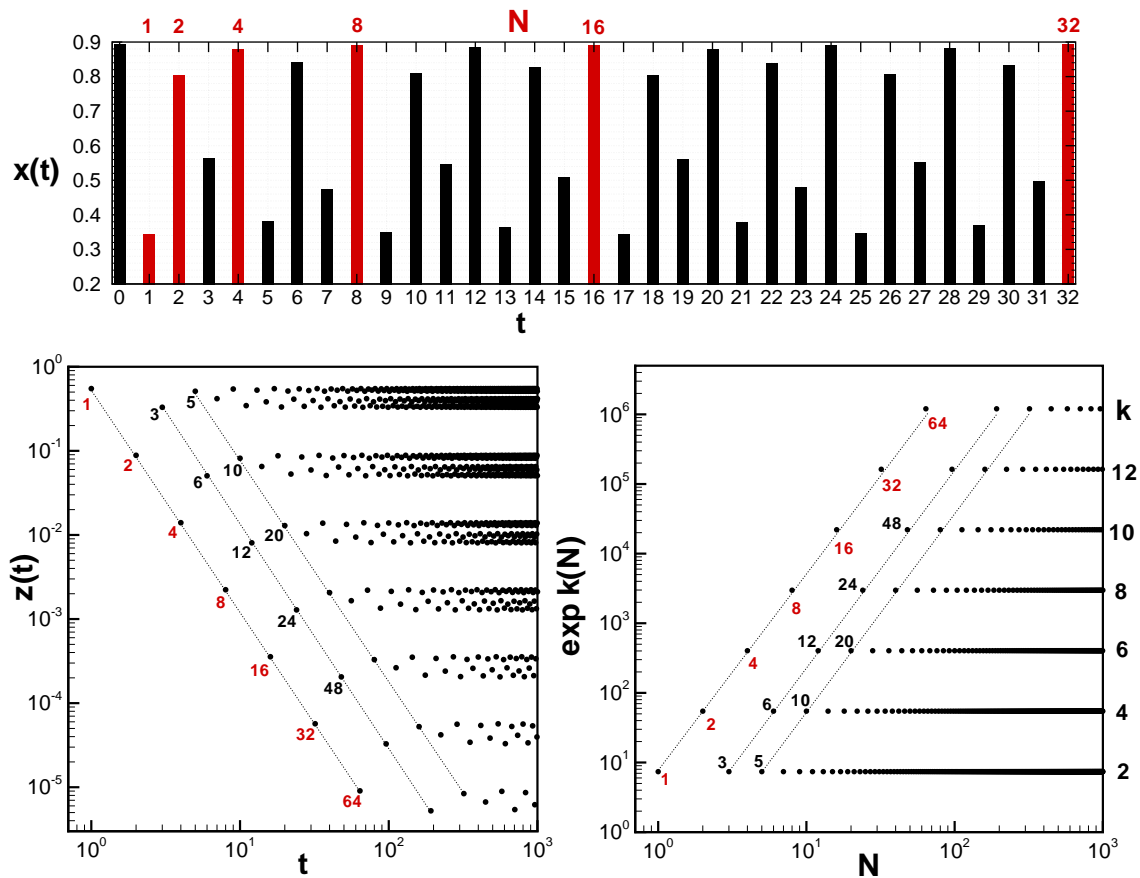}
\caption{\textit{Up:} Series $x(t)$ as a function of time $t$ for the first $10^{6}$ data generated from a logistic map at the period-doubling
accumulation point (only the first 33 data are shown). The data highlighted in red are associated with specific subsequences of nodes (see the text). \textit{Left:}
Log-log plot of the rescaled variable $z(t)=f(1/2)-x(t)$ as a function of $t$, for the same series as the upper panel. This rescaling
is performed to reflect the multifractal structure of the attractor \cite{baldovin1, mayoral1}. The order of visits to some specific data subsequences is highlighted.
\textit{Right:} Log-log plot of  $\exp k(N)$ as a
function of the node $N$ of the Feigenbaum graph generated from the same
time series as for the upper panel, where $N=t$. The distinctive band pattern of the attractor is recovered, although
in a simplified manner where the fine structure is replaced by single lines of constant degree. The order of visits to some specific node subsequences is highlighted (see the text).}
\label{log_k}
\end{figure*}

\noindent Consider the logistic map $x_{t+1}=f(x_t)=\mu x_t(1-x_t)$, $0\leq x\leq 1$, $0\leq
\mu \leq 4$, at the period-doubling accumulation point $\mu _{\infty}=3.5699456...$ and generate a trajectory $\{x_{t}\}_{t=0,1,2,...}$ within the attractor.
Since the position of the maximum of the map $x_{max}=1/2$ belongs to the attractor we can
choose $x_{0}=f(1/2)$ as the initial condition. (The position of the maximum of the unimodal map belongs to all the superstable periodic attractors or supercycles and the accumulation point of these attractors, the attractor at the transition to chaos, contains this position \cite {Feigenbaum}). A sample of the corresponding time series is shown in the top panel of figure \ref{log_k}, whereas in the left panel of the
same figure we represent in logarithmic scales the positions of the rescaled
variable $z_{t}=f(1/2)-x_{t}$. A similar rescaling was proposed in previous works to put in evidence the striped intertwined
self-similar structure of the trajectory positions as they are visited
sequentially, and reflects the multifractal structure of the attractor \cite{baldovin1, mayoral1}. In the right panel of the same figure we plot, in log-log scales, a representation of the relevant variable $\exp k(N)$, where $k(N)$ is the degree of node $N$ in the Feigenbaum graph associated to the original time series $x(t)$ (that is, $N\equiv t$). Notice that the distinctive band pattern of the attractor is recovered, although
in a simplified manner where the fine structure is replaced by single lines of constant degree. The HV algorithm transforms the multifractal attractor into a discrete set of connectivities, whose evolution mimics the intertwined fluctuations
of the map. It is convenient to write the node
index $N$ in the form $N\equiv m2^{j}$, where $j=0,1,2,...$ and $%
m=1,3,5,...$, so that running over the index $j$ with $m$ fixed selects
subsequences of data or nodes placed along lines with fixed slope. For illustrative purposes, the order of visits to either the associated data or nodes in some of these subsequences are explicitly shown in both panels of figure \ref{log_k}, as well as in the upper panel. Notice that along these subsequences, $\exp k(N)$ increase as a power law (i.e. subexponentially), and are associated to data with increasing larger values (upper panel). This is reminiscent of the data subsequences highlighted in the left panel, whose values $z(t)=f(1/2)-x(t)$ also show a monotonic power law decay.\\

\begin{figure}[tbh]
\centering
\includegraphics[width=0.45\textwidth]{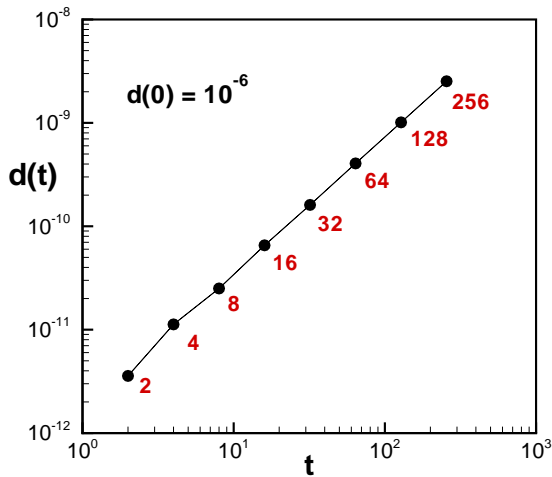}
\caption{Log-log plot of the distance between two nearby trajectories $d(t)=|x(t)-x'(t)|$ close to $x=1/2$, where $d(0)=10^{-6}$, measured at a precise time sequence $t=2^j, j=1,2,3,\dots$ (this path is also highlighted in the left panel of figure \ref{log_k}). Along this path trajectories evidence a subexponential expansion characterized by a power law separation. Note that both trajectories generate the same Feigenbaum graph, whose degree increases along this concrete path in the same vein as the trajectory separation: the network inherits the properties of trajectory expansion in the hierarchy of degrees (see right panel of figure \ref{log_k}). A similar behavior holds for trajectory separation along other paths.}
\label{expansion}
\end{figure}

From the direct observation of figure \ref{log_k} it is clear that all trajectory positions within each band in
the left panel (attractor) become nodes with the same degree in
the right panel (network). It is also worth noting at this point that there exists a degeneracy in the
positions of trajectories that are initiated off, but close to, the
attractor positions. In particular, if an ensemble of uniformly-distributed
initial positions is placed in a small interval around $x=1/2$ their
trajectories expand while remaining uniformly-distributed at later iteration times \cite{baldovin1, mayoral1} (see discussion below) and the HV algorithm assigns to all of them the same
Feigenbaum graph. This degeneracy accounts for the universal feature that
only a single Feigenbaum graph represents the transition to chaos for all
unimodal maps. It also determines the way in which fluctuations in
trajectory separations translate into their respective degree fluctuations: see figure \ref{expansion},
where we plot in logarithmic scales the distance of initially nearby trajectories for specific values of a time subsequence, and compare the expansion of this distance with the increment of the degree of the associated Feigenbaum graph along the same node subsequence, as highlighted in the right panel of figure \ref{log_k}. The properties of trajectory separation is thereby inherited by the graph and can be quantified through the increasing sequences of values for $\exp k(N)$.\\

Let us focus now in the Feigenbaum graph. The relabeling $N\equiv m2^{j}$ generates a one-to-one tiling of the
natural numbers. The aforementioned constant degree and fixed-slope lines in
the right panel of figure \ref{log_k} are, respectively, $k(N\equiv m2^{j})=2j+2$, \ $m=1,3,5,...$, $j$
fixed, and $k(N\equiv m2^{j})=2j+2$, $j$ $=0,1,2,...$, $m$ fixed (for example, the first path highlighted in red corresponds to $m=1, j=0,1,2,...$). The
scaling property associated with the former case is that all horizontal lines with
constant degree $k=2j+2$, $j$ fixed, can be overlapped into the bottom line $k=2$, $j=0$, via consecutive translations each consisting of a shift of $2$ in $k$ and a shift of $\log 2$ in $\log N$, that is, $k(N/2)=k(N)-2$, or alternatively, $\exp k(N/2)=\exp(-2)\cdot\exp k(N)$. This
scaling property has a parallel for the trajectory within the Feigenbaum
attractor where the shift of $2$ in $k$ is replaced by a shift of $2\log
\alpha $, where $\alpha $ is the absolute value of the Feigenbaum constant
\cite{baldovin1, mayoral1}. The scaling property associated with the latter
case is seen via the collapse of all fixed-slope lines of data into a single
sequence of values aligned along the first $m=1$ subsequence $k(2^{j})=2j+2$, $j=0,1,2,...$, when the numbers of nodes for each of the other
subsequences are rescaled consecutively by $N/m$, $m=3,5,7,...$.\\

\section{Fluctuating dynamics and graph-theoretical Lyapunov
exponents}
We recall that the standard Lyapunov exponent $\lambda $ accounts for the
degree of exponential separation of nearby trajectories, such that two
trajectories whose initial separation is $d(0)$ evolve with time distancing
from each other as $d(t)\sim d(0)\exp (\lambda t)$, $t\gg 1$, where $%
d(t)=|x_{t}-x_{t}^{\prime }|$. The sensitivity to initial conditions is $\xi
(t)=d(t)/d(0)$ and the general definition of the Lyapunov exponent is
\begin{equation}
\lambda =\lim_{t\rightarrow \infty }\frac{1}{t}\log \xi (t). \label{Lyapunov1}
\end{equation}%
Deterministic
fluctuations about $\lambda >0$ are permanently put out when $t\rightarrow
\infty $, but, as mentioned, at the period-doubling accumulation points of
unimodal maps $\lambda =0$ and trajectories display successive
subexponential separation and convergence for all $t$. Two possible
extensions of the standard Lyapunov exponent have been used when $\lambda =0$
while the amplitude of the fluctuations grows as a power law, or,
equivalently, expansion and contraction rates of trajectories are
logarithmic in time: Mori \textit{et} \textit{al} \cite{mori1} suggested the
expression
\begin{equation}
\lambda _{M}=\frac{1}{\log t}\log \xi (t).  \label{Lyapunov_Mori}
\end{equation}%
Although this expression captures subexponential fluctuation rates, this
quantity is not well defined if we want to measure the magnitude of rates
\emph{per unit time}, in which case we need a quantity that grows linearly
in time. This is important if we want to relate the growth of fluctuations
to some kind of entropic rates that derive from an extensive entropy
expression. In this respect a better-suited alternative for capturing
subexponential fluctuations is to deform the logarithm in equation \ref%
{Lyapunov1} by an amount that recovers the linear growth present for $\lambda
>0$. This is%
\begin{equation}
\lambda _{q}=\frac{1}{t}\log _{q}\xi (t),  \label{Lyapunov_Tsallis}
\end{equation}%
with $\log _{q}x\equiv (x^{1-q}-1)/(1-q)$ ($\log x$ is restored in the limit
$q\rightarrow 1$) where the extent of deformation $q$ is such that while the
expansion $\xi (t)$ is subexponential, $\log _{q}\xi (t)$ grows linearly
with $t$ \cite{baldovin1, mayoral1}. Notice that we have
eliminated the $t\rightarrow \infty $ limit in the definition of the
generalized Lyapunov exponents since fluctuations are present for all values of $%
t$ and the objective is to characterize them via a spectrum of generalized
exponents \cite{mori1, baldovin1, mayoral1}.

We define now a connectivity expansion rate for the Feigenbaum graph under
study. Since the graph is connected by construction (all nodes have degree $k\geq 2$), the uncertainty is only associated with a rescaled degree $k_{+}\equiv k-2$. To keep notation simple we make use of this variable and
drop the subindex $+$ from now on. The formal network analog of the
sensitivity to initial conditions $\xi (t)$ has as a natural definition $\xi
(N)\equiv \exp k(N)/\exp k(0)=\exp k(N)$, as suggested from figures \ref{log_k} and \ref{expansion}, where we are implicitly assuming that the expansion is always compared with the minimal one $\exp k(0)=1$ provided by nodes at positions $m2^0$ (i.e. with $k=0$).
Accordingly, the standard
network Lyapunov exponent is defined as
\begin{equation}
\lambda \equiv \lim_{N\rightarrow\infty }\frac{1}{N}\log \xi (N). \label{LyapunovN}
\end{equation}%
The value of $k(N)$ oscillates with $N$ (see
figure \ref{log_k}) but its bounds grow slower than $N$, as $\log N$, and
therefore in network context $\lambda =0$, in parallel to the ordinary
Lyapunov exponent at the onset of chaos. The logarithmic growth of the
bounds of $\log \xi (N)=k(N)$ is readily seen by writing $k(N=m2^{j})=2j$ as%
\begin{equation}
k(N)=\frac{2}{\log 2}\log \bigg(\frac{N}{m}\bigg).  \label{excessdegree1}
\end{equation}%
A first approach to study the subexponential fluctuations is to proceed
\emph{\`{a} la Mori} and define the following graph-theoretical generalized
exponent
\begin{equation}
\lambda _{M}=\frac{1}{\log N}\log \xi (N)=\frac{k(N)}{\log N}.  \label{Mori1}
\end{equation}%
To visualize the network growth paths $N=m2^{j}$, where $m=1,3,5,...$ is
fixed to a constant value and $j=1,2,3,...$, the expression above for $%
\lambda _{M}$ is written as%
\begin{equation}
\lambda _{M}=\frac{2j}{\log (m2^{j})}=\frac{2j}{\log m+j\log 2}.
\label{Mori2}
\end{equation}%
A constant $\lambda _{M}=2/\log 2$ for all $j$ is obtained only when $m=1$,
otherwise the same value is reached for all other $m$ when $j\rightarrow
\infty $. In general, a spectrum of Mori-like generalized exponents is
obtained by considering all paths $(m,j)$ such that $N\rightarrow \infty $.%
\newline

\noindent As indicated, the preceding generalization of the Lyapunov exponent is not
time extensive and it is therefore not useful if we ultimately wish to
relate the network link fluctuations to an entropy growth rate. With this
purpose in mind we deform the ordinary logarithm in $\log \xi (N)=k(N)$
into $\log _{q}\xi (N)$ by an amount $q>1$ such that $\log _{q}\xi (N)$
depends linearly in $N$, and define the associated generalized
graph-theoretical Lyapunov exponent as
\begin{equation}
\lambda _{q}=\frac{1}{\Delta N}\log _{q}\xi (N),  \label{Tsallis}
\end{equation}%
where $\Delta N=N-m$ is the lapse time between an initial condition $m2^0$ where $m$ is fixed (see figure \ref{log_k}), and position $N$. One obtains
\begin{equation}
\lambda _{q}=\frac{2}{m\log 2},  \label{lambdaq}
\end{equation}%
with $q=1-\log 2/2$ and $j>0$. Eq. (8) can be corroborated via use of $\xi (N)=\exp (2j)$ together with the identifications of the other quantities in it (note that $j=0$ implies $\Delta N=0$ for which $\xi (N)=\exp(0)=1$ and $\lambda _{q}$ is trivially undefined).
\noindent According to Eq. (9) a spectrum of exponents is spanned by running over the
values of $m$, in parallel with the spectrum of generalized exponents
previously found at the transition to chaos in unimodal maps \cite{mayoral1}%
, where the value of the parameter, $q=1-\log 2/2$ in the Feigenbaum graph
is to be compared with $q=1-\log 2/\log \alpha $ (where $\alpha$ is a Feigenbaum constant) in the unimodal map for
trajectories originating at the most compact region of the multifractal
attractor that are seen to expand at prescribed times when the least compact
region of the multifractal is visited \cite{mayoral1}.

\section{Entropic functionals and Pesin-like identities}
To complete our arguments we summon up the persistency property of
trajectory distributions of unimodal maps at the period-doubling onset of
chaos. That is, for a small interval of length $l_{1}$ with $\mathcal{N}$
uniformly-distributed initial conditions around the extremum of a
unimodal map (\textit{e.g.} $x=1/2$ for the logistic map we use), all
trajectories behave similarly, remain uniformly-distributed at later times, follow the concerted pattern shown in the left panel of Fig. 2. (see
\cite{baldovin1, mayoral1} for details) and expand subexponentially at prescribed time paths (see figure \ref{expansion}). Studies of entropy growth associated with an initial distribution of positions with iteration time $t$ of several chaotic maps \cite{Latora1} have established that a linear growth occurs during an intermediate stage in the evolution of the entropy, after an initial transient dependent on the initial distribution and before an asymptotic approach to a constant equilibrium value. At the period-doubling transition to chaos it was found \cite{baldovin1, mayoral1} that (i) there is no initial transient if the initial distribution is uniform and defined around a small interval of an attractor position, and (ii) the distribution remains uniform for an extended period of time due to the subexponential dynamics. We denote this distribution
by $\pi (t)=1/W(t)$ where $W(1)=l_1/\mathcal{N}$ is the number of cells that tile the initial interval $l_1$, and $l_{t}$ is the total length
of the interval that contain the trajectories at time $t$. Since iteration-time dependence transforms into node-value
dependence we inquire about how $\pi $ scales along the set of nodes $N=2^{j}
$ with $k=2j$ links, $j=0$, $1$, $2$,\ldots, that is, while $\pi$ is defined in the map its properties are reflected in the
network. After $t=2^j$ iterations, the loss of information associated with the fact that the same network is generated by all of the $\mathcal{N}$ initial conditions is represented in the network context by the expansion of the initial interval $l_1$ into $l_t=\exp k(2^j)=l_1 \exp (2j)$ after $t=2^j$.
Therefore we find that at time $t=2^j$ the network scaling properties prescribe a larger number of cells to tile the new interval $W(2^j)=W(1)\exp k(2^j)$. After normalization, we find that the $j$ dependence of $\pi$ reads


\begin{equation}
\pi(2^{j})=W_{j}^{-1}=\exp (-2j). \label{pi1}
\end{equation}%
That is, the uniform distributions $\pi $ for the consecutive
node-connectivity pairs (%
$2^{j},2j$) and ($2^{j+1},2(j+1)$) scale with the same factors noticed
above when discussing the right panel of figure 2. Namely, when $m=1$ these
pairs can be made equal by a shift of $\log 2$ in $\log N$ and a shift of $2$ in $k$.
By extension of the argument
the same expression holds for all other values of $m$. Since
\begin{equation}
W_{j}=\exp (2j)=\left( \frac{N}{m}\right) ^{2/\log 2},  \label{numberW1}
\end{equation}%
the ordinary entropy associated with $\pi $ grows logarithmically with the
number of nodes $N$, $S_{1}\left[ \pi (N)\right] =\log W_{j}\sim \log N$.
However, the $q$-deformed entropy
\begin{equation}
S_{q}\left[ \pi (j)\right] =\log _{q}W_{j}=\frac{1}{1-q}\left[ W_{j}^{1-q}-1%
\right] ,  \label{q-entropy1}
\end{equation}%
where the amount of deformation $q$ of the logarithm has the same value as
before, grows linearly with $N$, as $W_{j}$ can be rewritten as
\begin{equation}
W_{j}=\exp _{q}[\lambda _{q}\Delta N], \label{numberW2}
\end{equation}%
with $q=1-\log 2/2$ and $\lambda _{q}=2/(m\log 2)$. Therefore, if we define
the entropy growth rate
\begin{equation}
h_{q}\left[ \pi (N)\right] \equiv \frac{1}{N-m}S_{q}\left[ \pi (N)\right] \label{entropyrate2a}
\end{equation}%
we obtain
\begin{equation}
h_{q}\left[ \pi (N)\right] =\lambda _{q},  \label{entropyrate3}
\end{equation}%
a Pesin-like identity at the onset of chaos (effectively one identity for
each subsequence of node numbers given each by a value of $m=1,3,5,...$).\\
\section{Conclusions}
In conclusion, the transcription into a network of a special class of time
series, the trajectories associated with the attractor at the
period-doubling transition to chaos of unimodal maps, via the HV algorithm
has proved to be a valuable enterprise \cite{tolo1} as it has led to the
uncovering of a new property related to the Pesin identity in nonlinear
dynamics. The HV method leads to a self-similar network with a structure
illustrated by the related networks of periods $2^{n}$, $n=0,...,5$, shown
in Fig. \ref{grafos de feigenbaum1}. Under the HV algorithm many nearby
trajectory positions lead to the same network node and degree, all positions
within one band in the left panel of Fig. \ref{log_k} lead
to the same line of constant degree in the right panel of the figure. Only
when trajectory positions cross a gap between bands in the left panel the
corresponding node increases its degree by two new links. Also trajectories
off the attractor but close to it transform into the same network structure.
The degrees of the nodes span all even numbers $k=2j$, $j=0,1,2,...$, and we
have studied how these fluctuate as the number of nodes increases. The
fluctuations of the degree capture the core behavior of the fluctuations of the
sensitivity to initial conditions at the transition to chaos and they are universal
for all unimodal maps. The graph-theoretical
analogue of the sensitivity was identified as $\exp (k)$ while the amplitude of the variations of $k$ grows logarithmically with
the number of nodes $N$. These deterministic fluctuations are described by a discrete spectrum of generalized graph-theoretical
Lyapunov exponents that appear to relate to an equivalent spectrum of generalized entropy growths, yielding a set of
Pesin-like identities. The definitions of these quantities
involve a deformation of the ordinary logarithmic function that ensures their
linear growth with the number of nodes. Therefore the entropy expression involved is
extensive and of the Tsallis type with a precisely defined index $q$. A salient feature of the application of the HV algorithm to dynamical
systems time series is direct access to the degree distribution and
therefore to the entropy associated with it. Therefore this seems to be a
good tool to discover, or, inversely, to construct, network entropy properties
as studied here.
The capability of the
methodology employed in this study to reveal the occurrence of structural
elements of this nature in real networked systems that grow in time \cite{New03, New10}
appears to be viable.

\textbf{\emph{Acknowledgements}.}
We acknowledge financial support by the MEC and Comunidad de Madrid (Spain) through Project
Nos. FIS2009-13690 and S2009ESP-1691 (BL and LL), support from CONACyT \& DGAPA (PAPIIT IN100311)-UNAM (Mexican agencies) (AR), and
comments of anonymous referees.

\end{document}